\documentstyle[]{article}

\begin{document}
\begin{center} {\Large {\bf Bloch oscillations: atom optical
interpretation, realizations, and applications
}}\\[1cm]
Karl-Peter Marzlin
\footnote{e-mail: peter.marzlin@uni-konstanz.de}
and J\"urgen Audretsch
\footnote{e-mail: juergen.audretsch@uni-konstanz.de}
\\[2mm]
Fakult\"at f\"ur Physik
der Universit\"at Konstanz\\
Postfach 5560 M 674\\
D-78434 Konstanz, Germany
\end{center}
$ $\\[3mm]
\begin{minipage}{15cm}
\begin{abstract}
The cyclic motion of particles in a periodic potential under the
influence of a constant external force is analyzed
in an atom optical
approach based on Landau-Zener transitions between two resonant
states. The resulting complex picture of population transfers can
be interpreted in an intuitive diagrammatic way.
The model is also applied to genuine atom optical systems and
its applicability is discussed.
\end{abstract}
\end{minipage}
$ $ \\[1cm]
{\bf Introduction}\\
Bloch oscillations have been predicted by Bloch \cite{bloch29}
and Zener \cite{zener34} in connection with the electronic transport
in crystal lattices: If a constant static electric field is
acting on electrons moving in a periodic potential with period $L$,
the electron momentum turns out to be periodic in time with
frequency $\omega_B = e E L/\hbar$. Accordingly the driving electric
field $E$ leads to oscillations of the electrons. In normal
crystals these oscillations cannot be seen because of dephasing
collisions of the electrons. But in semiconductor superlattices the
accompagnying radiation has been observed \cite{waschke93}. The recent
experiments of Ben Dahan {\em et al.} \cite{salomprep} represent a
new step in the detection of Bloch oscillations.
They are for the
first time detected in the domain of atom optics. The
authors study atoms (instead of electrons) moving in a periodic
potential whereby the constant external force is simulated in
tuning linearly in time the two counterpropagating laser waves
which create the potential. Other atom optical systems too show
Bloch oscillations. Evidently the related oscillations in space
may be used to trap and store atoms.
An example based on a three-level atom has
been discussed by the authors in Ref. \cite{maau96}. On the
background of growing interest in Bloch oscillations of atoms
it seems to be desirable to give a genuine atom optical
interpretation of Bloch oscillations. We will show below that
in this approach
Landau-Zener transitions, which are known in connection with
level crossing and the related population transfer between the levels
of 2-level atoms, represent the dominating physical effect. Based
on this we study atom optical realizations of Bloch oscillations
and show how the respective systems can be used to drive
trapped atomic
clouds with almost constant velocity
thus forming an "atomic elevator". The range of application
where the model may be used is discussed.\\[3mm]
{\bf Bloch oscillations from an atom optical
point of view}\\
A typical problem in theoretical
solid state physics is the motion of
electrons in a periodic potential with constant driving force
being applied, in addition. This force $F$
may go back to a static
static electric field or a gravitational field.
In the one-dimensional case the Hamiltonian is given by
$H_s = p^2/(2M) - F x  + V(x)$
where the potential $V$ is periodic with lattice distance $L$,
$V(x+L) = V(x)$. If $V$ fulfills the Dirichlet conditions
(roughly, it can have only a finite number of finite jumps)
then $V$ can be written as
\begin{equation} V(x) = \frac{\alpha_0}{2} + \sum_{l=1}^\infty\Big
  \{ \alpha_l \cos (2\pi l x/L) + \beta_l \sin (2\pi l x/L)\Big \}
\label{v}\end{equation}
with real coefficients $\alpha_l$ and $\beta_l$. In momentum space
and in the interaction picture with respect to the acceleration
potential (i.e., taking $H_0 = - F x$)
the Schr\"odinger equation is given by
\begin{equation} i\hbar \partial_t \psi(\tilde{p},t) =
  \Big [\frac{(\tilde{p}+Ft)^2}{2M} + \frac{\alpha_0}{2}\Big ]
  \psi(\tilde{p},t) + {1\over 2}
  \sum_{l=1}^\infty \Big \{ \Omega^*_l \psi(\tilde{p}+ l\hbar k_0) +
  \Omega_l \psi(\tilde{p}-l \hbar k_o) \Big \} \end{equation}
where $k_0 := 2\pi /L$ is the border of the first Brioullin zone
and $ \Omega_l := \alpha_l + i \beta_l$. $\tilde{p}$ is a time
independent momentum parameter. $|\psi(\tilde{p},t)|^2$ is the
probability to measure at the time $t$ the momentum
$\tilde{p} + F t$ so that $\tilde{p}$ may
be interpreted as initial ($t=0$) momentum.
Obviously only a discrete ladder of states is coupled so that it is
favorable to introduce the notation
\begin{equation} \psi_n (t) := \psi (\tilde{p}_0 + n \hbar  k_0,t)
  \quad , \quad n= 0,\pm1 ,\pm 2, \ldots
\label{defn}\end{equation}
where $\tilde{p}_0$ is a momentum parameter. Setting
\begin{equation} \varepsilon_n := \frac{1}{\hbar} \Big \{
  \frac{(\tilde{p}_0 + n \hbar k_0 + Ft)^2}{2M} +
  \frac{\alpha_0}{2}\Big \} \label{epsn}\end{equation}
we obtain the dynamical equation
\begin{equation} i \partial_t \psi_n = \varepsilon_n \psi_n +
  {1\over 2} \sum_{l=1}^\infty \Big \{ \Omega^*_l \psi_{n+l} +
  \Omega_l \psi_{n-l}\Big \} \; . \end{equation}
which can be written in matrix form.
\begin{equation} i \partial_t
  \left ( \begin{array}{c} \vdots \\ \psi_{1} \\ \psi_0 \\
  \psi_{-1} \\ \psi_{-2} \\ \vdots \end{array} \right ) =
  \left ( \begin{array}{cccccc}  & \vdots & & &\vdots & \\ \cdots &
  \varepsilon_{1} & \Omega_1 & \Omega_2 & \Omega_3 & \cdots \\
  & \Omega_1^* & \varepsilon_0 & \Omega_1 & \Omega_2 & \\
  & \Omega_2^* & \Omega_1^* & \varepsilon_{-1} & \Omega_1 & \\
  \cdots & \Omega_3^* & \Omega_2^* & \Omega_1^* & \varepsilon_{-2}
  & \cdots \\ & \vdots & & & \vdots & \end{array} \right )
  \left ( \begin{array}{c} \vdots \\ \psi_{1} \\ \psi_0 \\
  \psi_{-1} \\ \psi_{-2} \\ \vdots \end{array} \right )
\label{matrix}\end{equation}

What types of transitions are caused by the matrix in Eq.
(\ref{matrix}) between the states $\psi_n$ and $\psi_{n+l}$,
for example?
To see this it is important to concentrate on resonances. The
states $\psi_n$ and $\psi_{n+l}$ are coupled by the $2\times 2$
matrix containing $\varepsilon_n$, $\Omega_l$, $\Omega_l^*$, and
$\varepsilon_{n+l}$. Because of Eq. (\ref{epsn}) the difference
between the related diagonal elements is given by
\begin{equation} \varepsilon_{n+l}-\varepsilon_n =
  \frac{l k_0}{M} \Big \{ {1\over 2}(2n + l)\hbar k_0 + \tilde{p}_0 +
  F t\Big \} \label{deleps}\end{equation}
Accordingly there will be a transition when this
difference becomes small (resonance case). It happens at a time
$t_{n,l} >0$ when an initial momentum  $\tilde{p}_0$ has been
changed by the influence of the force to a multiple of $\hbar k_0
/2$:
\begin{equation} \tilde{p}_0 + F t_{n,l} = -\hbar k_0 (n + l/2)
\; .\label{tdef}\end{equation}
The r.h.s. agrees with the momentum of the particle at the
beginning of the transition. To work out the time $t_{n,l}$, i.e.,
to fix $n$ and $l$ for the transition between two states $\psi_a$
and $\psi_b$ with $a>b$, one has to take into account that $n+l>n$.
Accordingly one can read off directly that $n$ is equal to the
smaller index ($b=n, l= a-b$).

We now assume that the recoil shift $\hbar k_0^2/(2M)$, which
determines the magnitude of the most important term (quadratic in
$l$) in Eq. (\ref{deleps}), is large compared to $|\Omega_l|$.
In this case all transitions between
$\psi_n$ or $\psi_{n+l}$ and any other state will be far off
resonance at the time $t_{n,l}$. Each individual transition (beside
$\Omega_l$) is then (in general)
highly suppressed and we will assume that
around $t= t_{n,l}$ the transition with $\Omega_l$ is the only one
which has to be taken into account for the pair of states
$\psi_n$ and $\psi_{n+l}$. Eq. (\ref{deleps}) shows that more than
two states cannot be in resonance with each other and
reveals the physical structure of the fundamental
transition involved. The energy difference (\ref{deleps})
approaches zero linearly in time. Accordingly the important
observation is that we have in resonance a {\em Landau-Zener
transition} (LZT) \cite{landau32,zener32}.

The efficiency of the LZT is determined by the
probability that the particles stay in the initial state
\begin{equation}
P_{\mbox{{\footnotesize stay}}}(l) = \exp \Big \{ -\pi
   \frac{|\Omega_l|^2}{4 |l k_0 a|} \Big \}
   \; . \label{lzerg} \end{equation}
This is an exact expression for the asymptotical transition
probability which is a good approximation if the time between
two subsequent transitions being in resonance is much larger than
the LZT time
\begin{equation} t_{L.Z.} := \frac{|\Omega_l|^2}{|l k_0
  a|^{3/2}} \label{tlz} \end{equation}
Characteristic for the LZT between $\psi_n$ and $\psi
_{n+l}$ is that the states in resonance are at equal kinetic
energy and that according to Eq. (\ref{defn}) there is a momentum
transfer of $\pm l\hbar k_0$ depending on the direction of the
transition.

Let us now describe the resulting
time development in detail starting with a
particular initial state. Parallel to the discussion of the
equations above we will sketch a more intuitive diagrammatic
picture to make the underlying physics transparent.  For this we
show in the figure the kinetic energy of the particles as a function
of their momentum $p(t)$. The horizontal arrows denoted by
$|\Omega_l|$ represent LZTs. The
respective transition probabilities
$P_{\mbox{{\footnotesize trans}}}
= 1 - P_{\mbox{{\footnotesize stay}}} $
are given by Eq. (\ref{lzerg}).
They fall off rapidly with increasing $l$ if the $\Omega_l$
do not grow with $l$.

With regard to later applications we assume that $F$ represents
the gravitational acceleration, $F = -M g $ with $g>0$. The x-Axis
points vertically upwards. Let us for simplicity assume that the
LZTs have vanishing duration. In this case they happen
at the times $t_{n,l}$ of Eq. (\ref{tdef}).
Between the LZTs the particles fall almost freely
(i.e., move along the parabola to the left).
We start at $t=0$ in the state $\psi_0$ with momentum $\tilde{p}_0$.
Transitions from $\psi_0$ are possible with $|\Omega_l|$ to
$\psi_{+l}$ at the time $t_{0,l}$ (compare the remark after Eq.
(\ref{tdef})) and to $\psi_{-l}$ at the time
$t_{-l,l}$ with momentum transfer $-l \hbar k_0$.
Let us assume for example
$+1/2 < \tilde{p}_0/(\hbar k_0) < 1$. Then it can be read off from
Eq. (\ref{tdef}) that the smallest $t_{0,l}$ or $t_{-l,l} > 0$ is
obtained for $l=+1, n=-1$. The corresponding transition is $\psi_0
\rightarrow \psi_{-1}$ with a momentum change $-\hbar k_0$,
compare Eqs. (\ref{defn}) and (\ref{matrix}). Turning to the figure
this means that we start with $p(t=0) = \tilde{p}_0$, move in free
fall along the parabola until at the time $t_{-1,1}$ the momentum
$p=\hbar k_0 /2$ is reached. Then the first LZT (in
this case with $|\Omega_1|$) may happen. The transition ends in the
state $\psi_{-1,1}$ with momentum $-\hbar k_0/2$. The
transition probability
depends on $P_{\mbox{{\footnotesize stay}}}$ and is therefore
determined by $|\Omega_1|$.
To follow the time development
the most transparent procedure is to start our clock anew. We
therefore take now as initial state
$\psi_0$ with $p=\tilde{p}_0 = -\hbar k_0/2$.
Then the smallest $t_{n,l}$ is because of
Eq. (\ref{tdef}) obtained for $n + l/2 = 1$ and accordingly for
$l=2$ and $n=0$. This is the transition
for which $\varepsilon_2 -\varepsilon_0$ in Eq. (\ref{deleps})
vanishes. Therefore we have $\psi_0 \rightarrow \psi_2$. The
related momentum transfer is $+2\hbar k_0$ so that we end with
momentum $p=\hbar k_0$. In the figure this corresponds to
the free fall from
$p=-\hbar k_0 /2$ to $p=-\hbar k_0$ and the $|\Omega_2|$
transition to the right with a probability obtained from Eq.
(\ref{lzerg}) with $l=2$ and $|\Omega_2|$. After this, those
particles which have made the transition fall again and so on.

The larger $l$ of $\Omega_l$ is, the larger is according to Eq.
(\ref{lzerg}) the possibility that the particles do not make the
$|\Omega_l|$ transition but follow the parabola to the left
(free fall). If $\Omega_l$ does not increase with $l$,
the $|\Omega_1|$ transition is the most effective one.
Accordingly, if particles start with $-1/2 < \tilde{p}_0/(\hbar
k_0) < +1/2$ they will fall until they reach $p(t)/(\hbar k_0) =
-1/2$ when, for small $P_{\mbox{{\footnotesize stay}}}$, they will
practically all be kicked to $p/(\hbar k_0) = +1/2$ from where they
fall again until $p(t)/(\hbar k_0) = -1/2$,
are kicked again and so on. These are the {\em simple Bloch
oscillations} with Bloch frequency $\omega_B = |F| L/\hbar$.
But as we have seen above there is the non vanishing probability
$P_{\mbox{{\footnotesize stay}}}(l=1)$ that the particles
carry on falling (Zener tunneling) and may then be subject to
{\em higher order Bloch oscillations} with LZTs
$|\Omega_2|, |\Omega_3|, \ldots$ in both directions if terms with
$l\neq 1$ in the potential $V(x)$ of Eq. (\ref{v}) don't vanish so
that a rather complex situation with many different
oscillations arises. For
$\Omega_{l\neq 1} =0$ the potential is of trigonometric shape and
the only oscillations which may happen for $\tilde{p}_0 >0$ are
simple Bloch oscillations. This completes our atom
optical approach in which the involved dynamics related to the
general periodic potential $V(x)$ of Eq. (\ref{v}) is based on
LZTs
in 2-level systems. \\[3mm]
{\bf Atom optical realizations}\\
Atoms with only a few energy levels are well studied objects of atom
optics. Many effective realizations can be found. It is therefore not
surprising that several experimental set ups can be proposed where
Bloch oscillations are the dominating
effect. We have made one proposal in
Ref. \cite{maau96} in connection with the gravito-optical trapping
of 3-level atoms. A 3-level $\Lambda$-system is exposed to two
counterpropagating laser fields (inducing Raman transitions). It is
closed by a magnetic hyperfine field tuned to be in resonance with
the transitions between the two ground states. The influence of a
homogeneous gravitational driving
field is taken into account. A discussion
in terms of dressed states leads to Eq. (34) of Ref. \cite{maau96}
which agrees with Eq. (\ref{matrix}) with $\Omega_l =0$ for $l\neq
1$. Again for appropriate initial conditions a sequence of up and
down motions is obtained which are simple Bloch oscillations.

Making again use of the Earth's gravitational acceleration $\vec{g}$
one may alternatively consider a 2-level atom in a resonant standing
laser wave with the Hamiltonian
\begin{equation}
H_{2\times 2} = {\bf 1}\Big \{ \frac{\vec{p}^2}{2M} -M \vec{g}
  \cdot \vec{x}\Big \} + \frac{\hbar \omega_0}{2} \sigma_3 -
  \Omega \cos (\vec{k}\cdot \vec{x})\cos (\omega_0 t) \sigma_1
\label{2niv}\end{equation}
where $\sigma_i$ are the Pauli matrices and $\omega_0$ is the
resonance frequency. Turning to dressed states by means of
the unitary transformation
\begin{equation} U^\prime = \left ( \begin{array}{cc} \exp [
  -i \omega_0 t/2 ]& \\ & \exp [i \omega_0 t/2] \end{array}
  \right )
  \exp [-i \pi \sigma_2 /4 ] \end{equation}
($|\psi^\prime \rangle = U |\psi \rangle$)
we find after the rotating wave approximation
\begin{equation} H^\prime = \{ \frac{\vec{p}^2}{2M} -M \vec{g}\cdot
  \vec{x}\} {\bf
  1} - {1\over 2} \hbar\Omega \cos (\vec{k}\cdot \vec{x}) \sigma_3
\; . \end{equation}
For each of the two states this is equivalent to the Hamiltonian
$H_s$ with $\Omega_l = 0$  for $l\neq 1$. This means that if the
Hamiltonian for the $\psi_l$ is written as a matrix that it is
tridiagonal. The Landau-Zener structure of the transitions is
particularly clear in this example.

An experimental realization of pure Bloch oscillations of
atoms has been discussed by Ben Dahan {\em et al.} \cite{salomprep}.
Here the atoms
are not subject to a potential $-M \vec{g}\cdot \vec{x}$
but are exposed to two
linearly chirped running laser waves.
That this is equivalent to the situation above can be seen
by applying the unitary transformation $\exp [i M \vec{g}\cdot
\vec{x}t /\hbar] \, \exp [-i \vec{p}\cdot \vec{g} t^2 /(2\hbar)]
\, \exp [-i M \vec{g}^2 t^3 /(6\hbar )]$ to Eq. (\ref{2niv}).
Physically this represents a transition to an accelerated reference
frame. The result of the transformation
is a Hamiltonian with two chirped running waves
\begin{equation} H= {\bf 1}\frac{\vec{p}^2}{2M}  + \frac{\hbar
  \omega_0}{2} \sigma_3 + \frac{\hbar \Omega}{2} \Big \{
  \cos (\omega_0 t + \vec{k}\cdot \vec{g} t^2 /2 +
  \vec{k}\cdot \vec{x}) + \cos (\omega_0 t - \vec{k}\cdot \vec{g} t^2
  /2 -\vec{k}\cdot \vec{x})\Big \} \sigma_1
\end{equation}
{\bf Handling atoms with an atomic elevator}\\
The atom optical system presented in Eq. (\ref{2niv})
allows a generalization
which is of interest for the manipulation of cold atomic clouds.
Imagine that the electromagnetic field is not a standing laser
wave but is
composed out of two counterpropagating running laser waves
with constant frequencies $\omega_0 \pm \Delta \omega$. By making a
Galilei transformation to a moving reference frame with velocity
$\vec{v} = \vec{k} \Delta \omega /k^2 $,  which amounts to
the unitary
transformation $U= \exp [i \vec{p}\cdot \vec{v}t/\hbar]$, we
essentially are back to the original Hamiltonian (\ref{2niv}).
Therefore the two-level atom performs Bloch oscillations
from the point of view of this moving frame.
But this means that in the original frame which is at rest
the atomic velocity oscillates around the velocity $v$
which was determined by the laser detuning $\Delta \omega$.
A limited extension in space for a moving atomic cloud can be
obtained if simple Bloch oscillations ($|\Omega_1|$ transitions in
the figure) are induced by appropriate initial conditions.
This effect can be employed to move atoms in a cloud up or down in the
Earth's gravitational field without a recognizable acceleration.
\\[3mm]
{\bf Range of application}\\
We now argue that the above model can only be applied to atoms
moving in electromagnetic fields with frequencies at least in the
optical regime if gravitation is the driving force. To do so, we
first note that the time $\Delta t$
between two subsequent LZTs must be much longer than
the time $t_{L.Z.}$ required to complete one individual LZT.
On the other hand, $P_{\mbox{{\footnotesize stay}}}$
must be much less than 1 in order to have an effective LZT.
These conditions lead to the inequality
\begin{equation} k a \ll \Omega^2 \ll \frac{\hbar k^{5/2}
   a^{1/2}}{2M} \end{equation}
which implies $k\gg (2M a^{1/2}/\hbar)^{2/3}$. Inserting for $a$
the Earth's acceleration setting $M=10^{-26}$ kg (light atoms)
one sees that $k$ must lie well above $10^6$ m$^{-1}$. Due to the
small mass of electrons the condition on $k$ is much less restrictive
in a crystal: $k_0$ needs only to be well above $10^3$ m$^{-1}$.
Since typical lattice spacings $L$ in crystals are less than
1 nm this is easily fulfilled by the electrons.

\newpage
{\bf Figure caption:}\\
A quantum particle moves in a periodic potential  with Fourier
coefficients $\Omega_l$ and periodicity length $L$
under the influence of a constant force $F$.
The kinetic energy $E$ as function of its momentum $p$ is shown
($k_0 = 2\pi /L$). The arrows represent Landau-Zener transitions
between resonant momentum states. The particle moves along the
parabola with $p(t)= \tilde{p}_0 + F t$ until
one of these states is
reached. Then there is a certain probability determined by
$l$ and $\Omega_l$ to stay on the parabola or to make a
transition to the resonant state and to follow the parabola
afterwards to a new resonance point where the next transition may
happen. For a particle with initial momentum $|p|<\hbar k_0/2$ the
$\Omega_1$ transition leads to simple Bloch oscillations of the
momentum with Bloch frequency $\omega_B = |F| L/\hbar$.
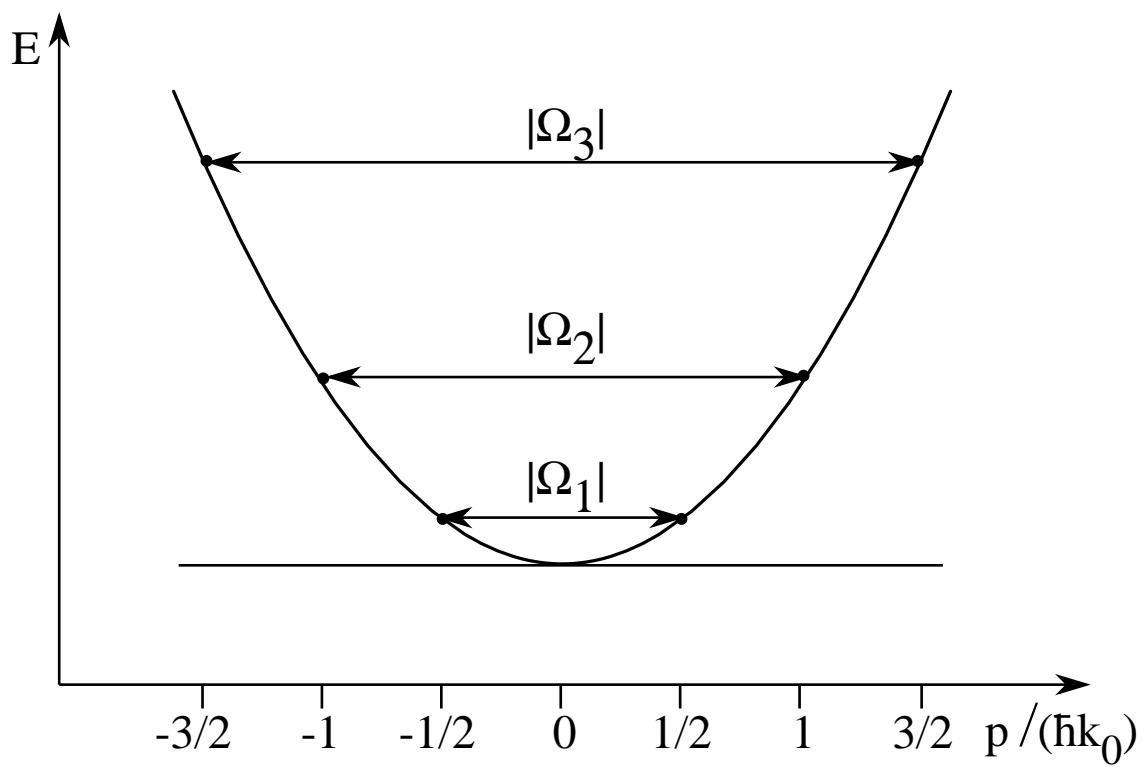
\end{document}